\documentclass[aps,pre,preprint,groupedaddress,floatfix]{revtex4-2}
\usepackage{graphicx} 
\usepackage{amsmath,amssymb,mathtools} 
\usepackage{bm}        
\usepackage{xcolor}    
\usepackage{siunitx}   
\usepackage{float}     
\usepackage{ulem}      
\let\emph\textit       
\makeatletter
\long\def\@makecaption#1#2{%
  \vskip\abovecaptionskip
  \begingroup
    \footnotesize
    \linespread{.9}\selectfont 
    \setlength{\parindent}{0pt}%
    \setlength{\parskip}{0pt}%
    \rightskip=0pt
    \leftskip=0pt
    \parfillskip=0pt
    \noindent
    \bfseries #1.\ \normalfont #2\par
  \endgroup
  \vskip\belowcaptionskip
}
\makeatother

\usepackage[left]{lineno}

\begin{document}

\title{
DNA end tethering through break-induced DNA--protein condensation
}

\author{Rakesh Das$^{1}$}
\email{rdas@pks.mpg.de}

\author{Tarun Mascarenhas$^{1}$}

\author{Nagaraja Chappidi$^{2,3}$}

\author{Simon Alberti$^{2,5}$}

\author{Frank J\"ulicher$^{1,4,5}$}
\email{julicher@pks.mpg.de}

\affiliation{$^{1}$Max Planck Institute for the Physics of Complex Systems, N\"othnitzer Stra{\ss}e 38, 01187 Dresden, Germany}

\affiliation{$^{2}$Biotechnology Center, Center for Molecular and Cellular Bioengineering, Technische Universit\"at Dresden, Tatzberg 47/49, 01307 Dresden, Germany}

\affiliation{$^{3}$Max Planck Institute of Molecular Cell Biology and Genetics, Pfotenhauerstra{\ss}e 108, 01307 Dresden, Germany}

\affiliation{$^{4}$Center for Systems Biology Dresden, Pfotenhauerstra{\ss}e 108, 01307 Dresden, Germany}

\affiliation{$^{5}$Cluster of Excellence Physics of Life, Technische Universit\"at Dresden,, 01397 Dresden, Germany}

\date{\today}

\begin{abstract}
Cells deploy robust mechanisms to repair DNA damage, safeguarding genomic stability and cellular health, but the physical principles underlying these processes remain incompletely understood.
Experiments show \emph{in vitro} that upon a DNA double-strand break, a DNA--protein condensate can tether the broken DNA ends before they disperse away, a critical step for subsequent repair biochemistry. 
However, it remains puzzling how such condensation reliably achieves spatiotemporal localization at the break site and captures both broken ends despite intrinsic stochasticity. 
Here, we propose that broken DNA ends can trigger a conversion of proteins from a soluble state to a condensate-competent state. 
Combining this idea with Brownian dynamics simulations and theory, we propose a physical mechanism for reliable DNA-end tethering. 
Simulations show that such break-induced conversion can drive local DNA--protein condensation with two possible outcomes: successful or failed tethering. 
To rationalize this, we construct an effective free energy landscape, identify the corresponding stationary states, and demonstrate that tethering is governed by a kinetic competition between polymer relaxation and condensation dynamics. 
Together, our study shows that DNA end-dependent conversion, coupled with DNA--protein condensation, can reliably tether broken DNA ends. 
\end{abstract}


\maketitle

\clearpage
\noindent
DNA, the genetic material in living systems, is under constant assault from endogenous and exogenous sources that induce various kinds of DNA damage \cite{PfeiferNatRevGenet2024, ElledgeMolCell2010, ScullyNatRevMolCellBiol2019}. 
Cells have robust mechanisms to detect and repair such damage. 
Upon damage, the biochemical responders assemble to form a repair site in the damage vicinity to eventually repair it, ensuring genomic integrity. 
The protein PARP1  is an early responder to DNA damage that is known to recognize the damage site, initiate repair pathways, and contribute to the repair process through PARylation\textemdash an enzymatically active process that synthesizes polymeric substrates enabling the local binding of downstream repair factors \cite{AlbertiCell2024, AleksandrovMolCell2018, RayChaudhuriNatRevMolCellBiol2017, PascalCurrOpinStructBiol}.
Misregulation in these damage responses is associated with  cellular pathology \cite{HoeijmakersNature2021, ShimizuCellMetab2014, TiwariAmJHumGenet2019, NiedernhofereLife2021}.

In order for these complex repair processes to be successful, first it is essential to keep the two broken DNA ends in physical proximity after the break. 
Recent work  has revealed the biochemical orchestration underlying the cellular response to DNA double-strand breaks \cite{AlbertiCell2024}. 
In this work, the biophysical relevance of DNA--PARP1 co-condensation was established in repair site assembly.
Furthermore, they showed \emph{in vitro} that the DNA--PARP1 co-condensation at the site of DNA break could keep the broken ends together. 
However, it remains unclear how this DNA--protein condensation can occur rapidly at the right position to achieve tethering of the broken ends.

In principle, the physics of phase separation provides a general framework to account for this type of condensation \cite{ChaikinBook, JulicherAnnuRevCellDevBiol2014, JulicherAnnuRevCondensMatterPhys2024}.
DNA--protein condensation has been studied previously in other contexts \cite{JulicherNatPhys2021, BrangwynneCell2018, GrillNatPhys2022, GrillPNAS2022}.
For example, pioneer transcription factors have been shown to form condensates on DNA \cite{GrillNatPhys2022}, and DNA--protein condensation has been demonstrated to generate mechanical forces \cite{JulicherNatPhys2021, GrillPNAS2022}.
Condensation of proteins \emph{in vitro} with DNA involves capillary forces \cite{DeGennesCapillarityBook}, which pull the polymer into the condensate. 
The steady-state size of the condensate is determined by a balance between these capillary forces and the opposing polymeric elasticity \cite{SiggiaScience1994}.
A strongly extended polymer does not support condensate formation, because the energy of condensation is not sufficient to further stretch the polymer.
In contrast, for a weakly extended polymer, condensates can nucleate at stochastic locations and grow, eventually coarsening into a single condensate. 
The physics of DNA--protein condensation is well understood. 
However, it remains unclear how DNA--protein condensation could tether broken DNA ends.
Furthermore, despite a high abundance of PARP1 in both the cell and \emph{in vitro} contexts \cite{AlbertiCell2024}, which in principle permits condensation at different locations, it is puzzling that a condensate appears precisely at the damage site.

Here, we address the question of how a DNA double-strand break could initiate nucleation quickly at the broken ends, such that a condensate forms that tethers the two broken ends.
We use Brownian dynamics simulations and the theory of DNA--protein condensation to discuss the minimal system that would be required to achieve tethering, as well as the conditions and probabilities of tethering.

In order to develop a simple model of break-induced condensation, we first highlight key experimental observations.
Studying PARP1 co-condensation with single DNA molecules reveals that condensates form only after DNA breakage, specifically at the broken DNA ends, and that these condensates can tether the broken ends together \cite{AlbertiCell2024}.
A separate bulk assay \cite{AlbertiCell2024} containing fragmented double-stranded DNA (dsDNA) and PARP1 in solution suggested a biochemical basis of such localized condensation.
While robust DNA–PARP1 co-condensation can occur under these conditions, it is markedly reduced when the ends of the dsDNA fragments are blocked with streptavidin. 
This  shows  that free dsDNA ends are key for the  DNA–PARP1 condensation observed in the single-DNA-molecule assay described above. 
We therefore propose that the PARP1 protein exists either in a soluble state $A$ or in a condensation-competent state $B$, and broken DNA ends locally facilitate the conversion of $A$ into $B$. 
The latter point implies that the broken ends act like catalysts, enabling a conversion that is not possible in the absence of a broken end.
We do not know the molecular details of the trigger of the condensation. 
For simplicity, we use a coarse grained model where proteins undergo a conversion from a state A to B to   represent enhanced condensation competence near the broken ends. 

\begin{figure*}[h!]
    \centering
    \includegraphics[width=0.95\textwidth]{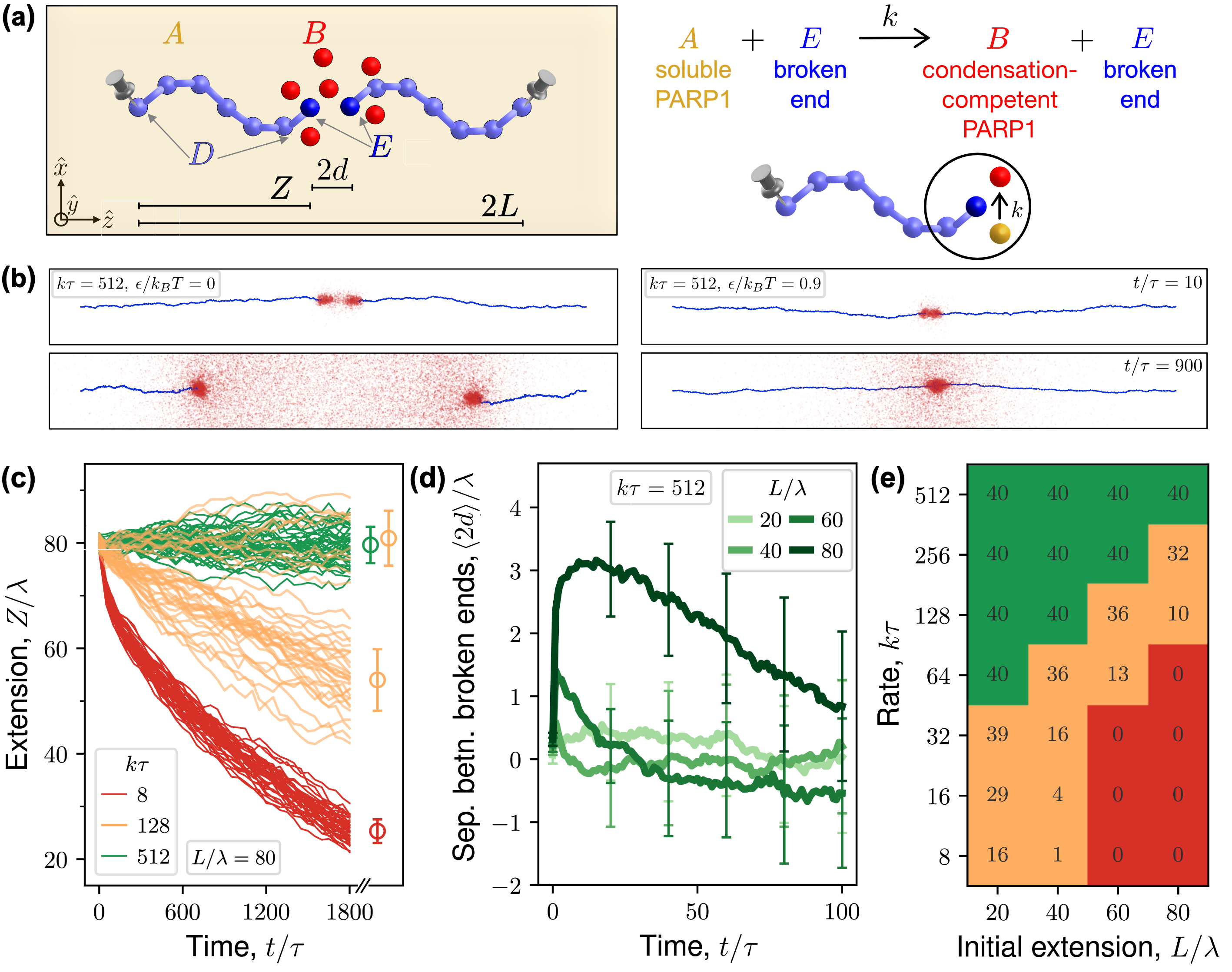} 
    \caption{
    {\bf Reliable tethering of broken polymer ends can be achieved through break-induced condensation.}
    {\bf a.} 
    Schematic representation of the break-induced condensation model. 
    Before the break, an intact bead-and-spring polymer was pinned in the medium by its terminal beads.
    Following a symmetric break in its backbone, two broken chains appear, each having a broken end bead $E$ (dark-blue) and other $D$ beads (blue). 
    Each $E$ locally triggers the conversion of soluble protein $A$ (golden, homogeneously dissolved in the background medium) into $B$ beads (red, representing the condensation-competent state of the protein).
    The system geometry is a rectangular parallelepiped, with the Cartesian unit vectors indicated. 
    Along $\hat z$, the separations between the pinned terminal beads of the polymer, the left pinned terminal bead and the corresponding broken end bead, and the two broken ends are $2L$, $Z$, and $2d$, respectively.
    Conversion rate is $k$.
    {\bf b.} 
    The system evolves following Brownian dynamics, with the terminal polymeric beads remaining pinned.  
    Typical simulation snapshots are shown for $L/\lambda = 80$ and several time $t$ after the break. 
    In the right panels, interactions between the polymer and $B$ beads drive polymer--$B$ condensate formation that tether the broken ends together. 
    No condensation or tethering occurs in the left panels. 
    See the main text for discussion on the similar spatial distribution of $B$ in both cases.
    {\bf c.} 
    For each $k$, $Z(t)$ are shown for $40$ independent simulation runs.  
    Late-time values of $Z$ indicate successful tethering for the full population (green) or only a partial population (orange), or tethering failure for the full population (red). 
    At the right margin, the mean $\pm$ s.d. for the respective populations at $t/\tau = 1800$ are shown.
    {\bf d.} 
    For each $L$, the mean $\pm$ s.d. over $40$ simulation runs are shown. 
    Despite early-time growth in $\langle d \rangle$ for $L/\lambda=80$, later dynamics leads to successful tethering.  
    {\bf e.} 
    Statistics of observed tethering outcomes for several values of $L$ and $k$. 
    Green, orange, and red colors indicate the full-population success, the partial population success, or full-population failure in tethering.
    Each numeric entry indicates the number of successful tethering events out of 40 independent simulation runs. 
    }
    \label{Fig_model}
\end{figure*}

In our Brownian dynamics simulations of break-induced condensation, we consider a bead-and-spring polymer embedded in a viscous medium. 
The medium may also contain individual $B$ beads, which can co-condense with the polymer. 
The consecutive beads in the polymer are labeled by $n = 1, \dots, 2N$. 
If there are $M$ number of $B$ beads, we label them by $n = 2N+1, \dots, 2N+M$.
Each polymeric bead represents a coarse-grained DNA segment with a size larger than its persistence length and belongs to species $D$, distinguishing it from the non-polymeric beads.
Bead $n$ is specified by its positions $\bm{x}_n$ and species label $\xi_n \in \{D, E, B\}$, for $n \in [1, 2N+M]$.
The two terminal $D$ beads, $n=1$ and $2N$, are pinned in space at a separation $2L$ along the $\hat{z}$-direction (Fig.~\ref{Fig_model}a).
At time $t=0$, with no $B$ beads in the medium ($M=0$), the polymer is symmetrically broken into two chains with $n = 1, \ldots, N$ and $n = N+1, \ldots, 2N$, respectively.
This creates two broken ends at $n=N$ and $n=N+1$ with species label $E$ (formerly $D$).
Bead positions evolve according to an overdamped Langevin equation with thermal noise:
\begin{eqnarray}
    \frac{\partial \bm{x}_n}{\partial t} = - \Gamma \frac{\partial H}{\partial \bm{x}_n} + \sqrt{2 \Gamma k_B T} \bm{\zeta}_n \quad.
    \label{EqBD}
\end{eqnarray}
Here, $\Gamma$ denotes the mobility, $k_B$ is the Boltzmann constant, $T$ the temperature. 
The stochastic term $\bm{\zeta}_n$ represents Gaussian white noise with zero mean and delta correlations, characterized by
$\langle \bm{\zeta}_n(t) \, \bm{\zeta}_m(t') \rangle = \bm{I} \, \delta_{nm} \, \delta(t-t')$,
where $\bm{I}$ is the identity tensor in three dimensions.
The $E$ beads, which represent the broken ends, catalyze the conversion of the soluble state $A$ to the condensation-competent state $B$, with rate $k$ (Fig.~\ref{Fig_model}a):
\begin{equation}
    A + E \xrightarrow{k} B + E \quad.
    \label{EqCatalysis}
\end{equation}
Here, we consider $B$ to be energetically favorable compared to the metastable state $A$, implying that the reaction is spontaneous.
For computational convenience, the state $A$ is  not explicitly represented, but considered as a homogeneous background field. 
$B$ beads are introduced with  a Poisson rate $k$ at a random position in proximity to an $E$ bead only if sufficient space is available (see Methods). 
Stochastic times for these attempts are generated using Gillespie algorithm.
We use a timescale $\tau$, a lengthscale $\lambda$, and the energyscale $k_B T$ as simulation units.
The interaction potential in eq.~\ref{EqBD} can be written as 
\begin{eqnarray}
    H = \sum_{n} \sum_{m > n} \epsilon_{nm} u(r_{n,m}) + \sum_{n=1,2N} h_{\mathrm{pin}}(\bm{x}_n) 
      + \sum_{n=1}^{N-1} h_{\mathrm{FENE}}(r_{n,n+1})  
      +  \sum_{n=N+1}^{2N-1} h_{\mathrm{FENE}}(r_{n,n+1}) \quad. \quad
\end{eqnarray}
Here, $\epsilon_{nm}$ depends on the species labels of beads $n$ and $m$, and takes three possible values:
(i)   $\epsilon_{nm} = \epsilon$ when $\xi_n \in \{D,E\}$ and $\xi_m = B$,
(ii)  $\epsilon_{nm} = \epsilon_1$ when $\xi_n, \xi_m \in \{D,E\}$, and 
(iii) $\epsilon_{nm} = \epsilon_2$ when $\xi_n, \xi_m = B$.
Beads separated at distance $r_{n,m} = | \bm{x}_n - \bm{x}_m |$ interact via a truncated Lennard-Jones $8$-$4$ potential specified by $u(r_{n,m})$.  
We set the interaction strength $\epsilon$ to a value that can drive polymer--$B$ co-condensation.  
The strengths $\epsilon_1$ and $\epsilon_2$ are chosen such that neither the polymer nor the $B$ beads can condense on their own.
The two terminal $D$ beads of the polymer are pinned to fixed positions by the harmonic potentials $h_{\mathrm{pin}}$.
The backbone connectivity of the consecutive polymeric beads in both chains are maintained via the finitely-extensible-nonlinear-elastic (FENE) spring potentials $h_{\mathrm{FENE}}$. 
See Methods for further details of our simulation system.

In the presence of catalysis ($k>0$), $B$ beads are generated in our simulations from the background $A$-field near the broken ends (see Fig.~\ref{Fig_model}b, Supplemental Movies~1--3 \cite{SM}). 
For sufficiently large $\epsilon > 0$, we observe that the $B$ beads condense together with the polymer to form a dense polymer--$B$ phase.
This phase separation phenomenon occurs close to the broken ends, and the resulting condensate can keep the broken ends together even at late times.
In contrast, despite the presence of $B$ beads, no tethering is observed in the absence of phase separation ($\epsilon=0$; also see Extended~Data~Fig.~\ref{Extended_Data_Fig1}).
These results highlight that (i) break-induced catalysis ($k>0$) provides a spatiotemporal regulation of $B$ production after the break and near the broken ends, and that (ii) phase separation ($\epsilon>0$) is required to form condensates and to tether the broken ends. 

Comparing the cases with and without phase separation in Fig.~\ref{Fig_model}b, we observe a similar localized distribution of $B$ near the broken ends in both situations.
This is because catalysis by $E$ drives the local conversion of $A$ into $B$, therefore $E$ acts as a source of new $B$ giving rise to a localized density profile (see Supplemental Material \cite{SM}). 
However, tethering of the broken ends is achieved only if phase separation occurs and polymer--$B$ condensates form.
In the following, we focus on this phase-separating case. 
Within this setting, $k$ controls the speed of condensation by determining the total amount of $B$ produced over time, and $\epsilon$ controls the segregation strength of condensation.

\begin{figure*}[h!]
    \centering
    \includegraphics[width=0.95\textwidth]{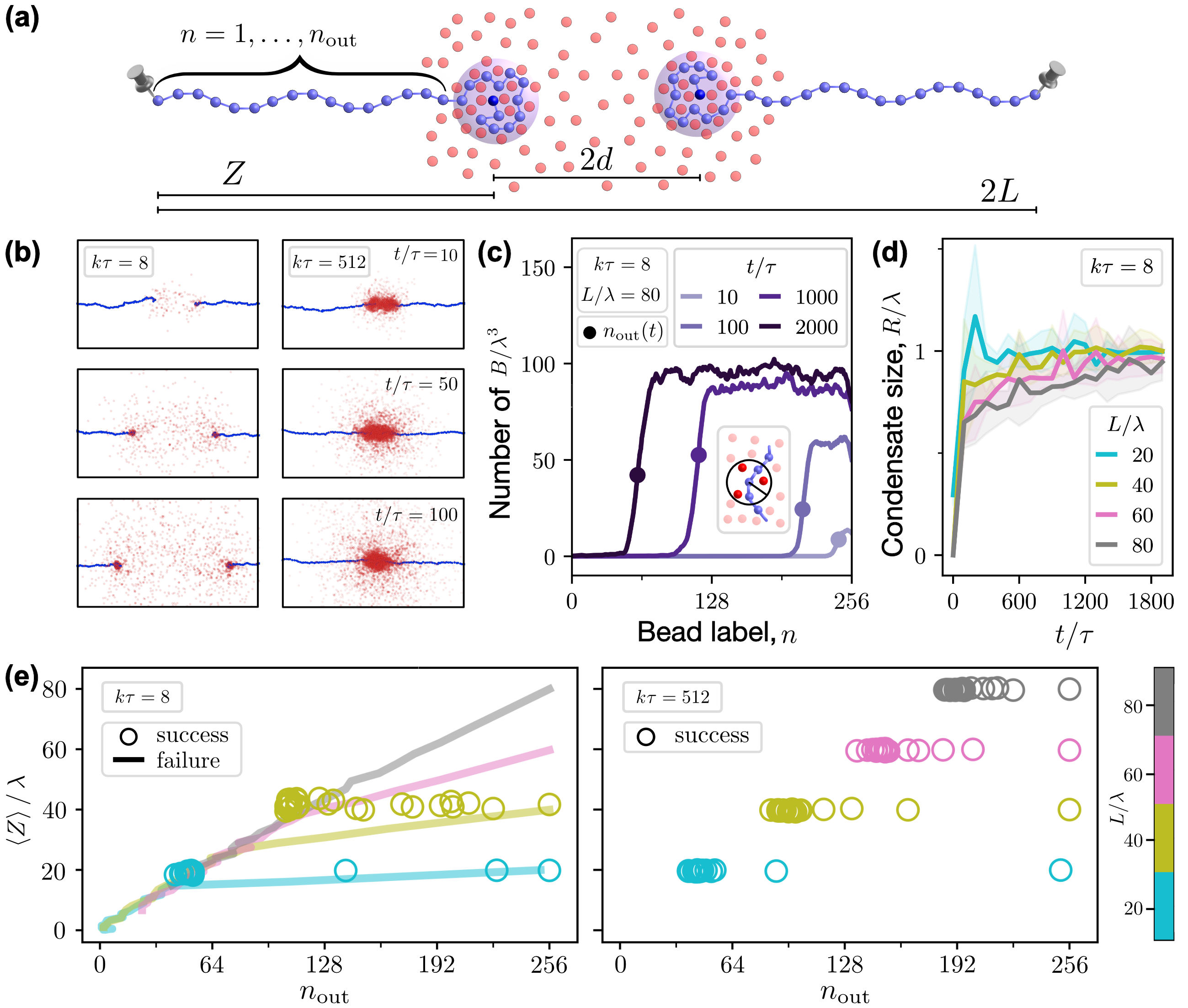} 
    \caption{
    {\bf Interplay between polymer dynamics and broken-end-dependent condensation governs tethering outcome.}
    {\bf a.} 
    Schematic of a typical system configuration, showing the broken chains ($D$ and $E$ beads in blue and dark-blue, respectively) and $B$ beads (red). 
    Condensates form and grow at the broken ends. 
    At a given $t$, the beads with labels $n = 1, \ldots, n_{\mathrm{out}}(t)$ in the broken chain lie outside the condensate, while the remaining beads of that chain, $n = n_{\mathrm{out}}(t)+1, \ldots, N$, are inside the polymer--$B$ condensate. 
    See the caption of Fig.~\ref{Fig_model}a for definitions of $L$, $Z$, and $d$.
    {\bf b.}
    Typical early-time simulation snapshots are shown for $L/\lambda = 80$, zoomed near the break site. 
    The growing condensates at the broken ends cannot fuse because the broken chains retract rapidly compared to the condensation dynamics for $k\tau = 8$. 
    In contrast, faster condensation for $k\tau = 512$ allows fusion of the condensates, thereby leading to successful tethering. 
    {\bf c.}
    Average number of $B$ within a unit spherical volume centered on polymeric bead $n$ is shown (see the schematic inside the panel). 
    For a given $t$, the data are averaged over $40$ simulation runs. 
    The plateau highlights the chain segments inside the polymer--$B$ condensate. 
    The $B$ density changes most sharply at $n = n_{\mathrm{out}}(t)$.
    {\bf d.} 
    The radius of gyration, computed from the positions of polymeric beads $n = n_{\mathrm{out}}(t)+1, \ldots, N$, is presented. 
    For each $L$, the mean $\pm$ s.d. are shown, averaged over $40$ simulation runs.
    {\bf e.} 
    System evolution in the $(n_{\mathrm{out}}, \langle Z \rangle)$ state space.  
    For each set of $(k, L)$ parameters, the ensemble consists of $40$ simulation runs, and averages are computed accounting for whether each run resulted in successful or failed tethering.
    Each ensemble-averaged trajectory originates from $n_{\mathrm{out}} = N$ ($N = 256$ here) and $\langle Z \rangle = L$ at $t = 0$, and shown for time $t \leq 2500\,\tau$. 
    At late times, a trajectory either reaches a steady state with $n_{\mathrm{out}} > 0$ and $\langle Z \rangle > 0$ (symbols, corresponding to tethering success), or approaches the complete incorporation of the broken chain into the condensate, $n_{\mathrm{out}} = 0$ (lines, corresponding to tethering failure). 
    The fluctuations observed in the trajectory corresponding to the tethering-success case for $k\tau = 8$, $L/\lambda = 40$ arise from the small population size (see Fig.~\ref{Fig_model}e for corresponding number).
    }
    \label{Fig_simu_trajectory}
\end{figure*}

To characterize the tethering, we define the extension of the broken chain by $Z = \left( \bm{x}_{N} - \bm{x}_{1} \right) \cdot \hat{z}$. 
The separation between the broken ends is defined by $2d = \left( \bm{x}_{N+1} - \bm{x}_{N} \right) \cdot \hat{z}$. 
The averages of both are related by $\langle d \rangle = L - \langle Z \rangle$, where $\langle \cdot \rangle$ indicates an ensemble average at a given time. 
We start our simulations from the initial condition $\langle Z \rangle = L$ corresponding to an intact polymer (see Methods), and at $t=0$ remove the connectivity between beads $N$ and $N+1$, thereby introducing the break. 
Two different outcomes are possible in each simulation: tethering success corresponding to an extended chain with $Z >0$ at long times, and failure with $Z$ approaching zero over time.
An example of repeated simulations is shown in Fig.~\ref{Fig_model}c for $L/\lambda = 80$.
For large catalysis rate, all simulations result in successful tethering (Fig.~\ref{Fig_model}c, green). 
At early times, the separation between the broken ends increases rapidly as they move apart, but then decreases as tethering sets in (Fig.~\ref{Fig_model}d).
For small catalysis rate, all simulations result in  failure (Fig.~\ref{Fig_model}c, red). 
In an intermediate case, the outcome is stochastic with partial success (Fig.~\ref{Fig_model}c, orange).
We further study the effect of the initial extension $L$ on the system dynamics. 
For shorter $L$ and large catalysis rate, the broken ends move apart slowly as the polymer is under lower tension (Fig.~\ref{Fig_model}d). 
Varying both catalysis rate and initial extension, the tethering success is most reliable for smaller initial extension and faster catalysis, while it typically results in failure for larger initial extension and slower catalysis (Fig.~\ref{Fig_model}e).
We can identify three physical processes governing these simulations:
(i) phase separation, giving rise to polymer--$B$ condensates at the broken ends, growing with time,
(ii) polymer relaxation of the chain segments outside the condensates, and
(iii) fusion of condensates. 
The tethering outcome is eventually governed by a kinetic competition between phase separation and polymer relaxation, which determines whether the condensates fuse (Figs.~\ref{Fig_simu_trajectory}a--b). 

Next, we study how this kinetic competition couples to the tethering outcome\textemdash success or failure. 
For this, first we characterize temporal progress of the condensation by identifying the polymeric bead $n_{\mathrm{out}}(t)$ at the  boundary separating the chain segments outside and inside the condensate, with $n=1, \ldots, n_{\mathrm{out}}$ and $n=n_{\mathrm{out}}+1, \ldots, N$,  respectively (Fig.~\ref{Fig_simu_trajectory}a).
Fig.~\ref{Fig_simu_trajectory}c shows the density of $B$ as a function of bead label $n = 1, \ldots, N$, and at $n = n_{\mathrm{out}}$ the density changes sharply from inside the condensate to the outside.
The radius of gyration $R$, calculated from the bead positions ${\bm{x}_n}$ for $n=n_{\mathrm{out}}+1, \ldots, N$, represents the size of the condensate growing at the broken end. 
Examining the case without fusion, we observe rapid post-break condensation, with $R$ quickly reaching a plateau (Fig.~\ref{Fig_simu_trajectory}d).

Fig.~\ref{Fig_simu_trajectory}e shows examples of the system dynamics in the state space spanned by $n_{\mathrm{out}}$ and $\langle Z \rangle$.
For each set of parameters $k$ and $L$, the trajectories are first classified according to their tethering outcome (success or failure, when both occur) within the full ensemble, and the shown trajectories are then obtained by averaging separately over these subclasses. 
Such ensemble trajectories evolve from the initial state $n_{\mathrm{out}} = N$ and $\langle Z \rangle = L$ at $t=0$, and are shown in Fig.~\ref{Fig_simu_trajectory}e for small and large catalysis rates (left and right panels, respectively).
For visual clarity, trajectories corresponding to tethering success are shown as symbols, while those corresponding to tethering failure are shown as lines. 
For the large catalysis rate, the fast growth of the condensates at the broken ends outpaces the retraction of the broken chains, allowing the condensates to fuse and thereby leading to tethering success. 
Consequently, the trajectories eventually reach steady states with $n_{\mathrm{out}} > 0$ and $\langle Z \rangle > 0$ (Fig.~\ref{Fig_simu_trajectory}e, right panel).
In contrast, for the small catalysis rate and large initial extension $L$, the broken chains retract so quickly that, despite rapid condensation at the broken ends, the condensates cannot fuse, resulting in tethering failure. 
Accordingly, the trajectories evolve toward $n_{\mathrm{out}} \to 0$ and $\langle Z \rangle \to 0$ over time (Fig.~\ref{Fig_simu_trajectory}e, left panel, for $L/\lambda = 60$ and $80$).
However, for small initial extensions, stochastic effects become important in determining the tethering outcome, giving rise to both types of trajectories, where symbols and lines both appear for the same parameter set (Fig.~\ref{Fig_simu_trajectory}e, left panel, for $L/\lambda = 20$ and $40$).
Our simulations reveal an interplay between the condensation dynamics and chain retraction giving rise to regimes where tethering success is either reliable or stochastic.

\begin{figure*}[h!]
    \centering
    \includegraphics[width=0.95\textwidth]{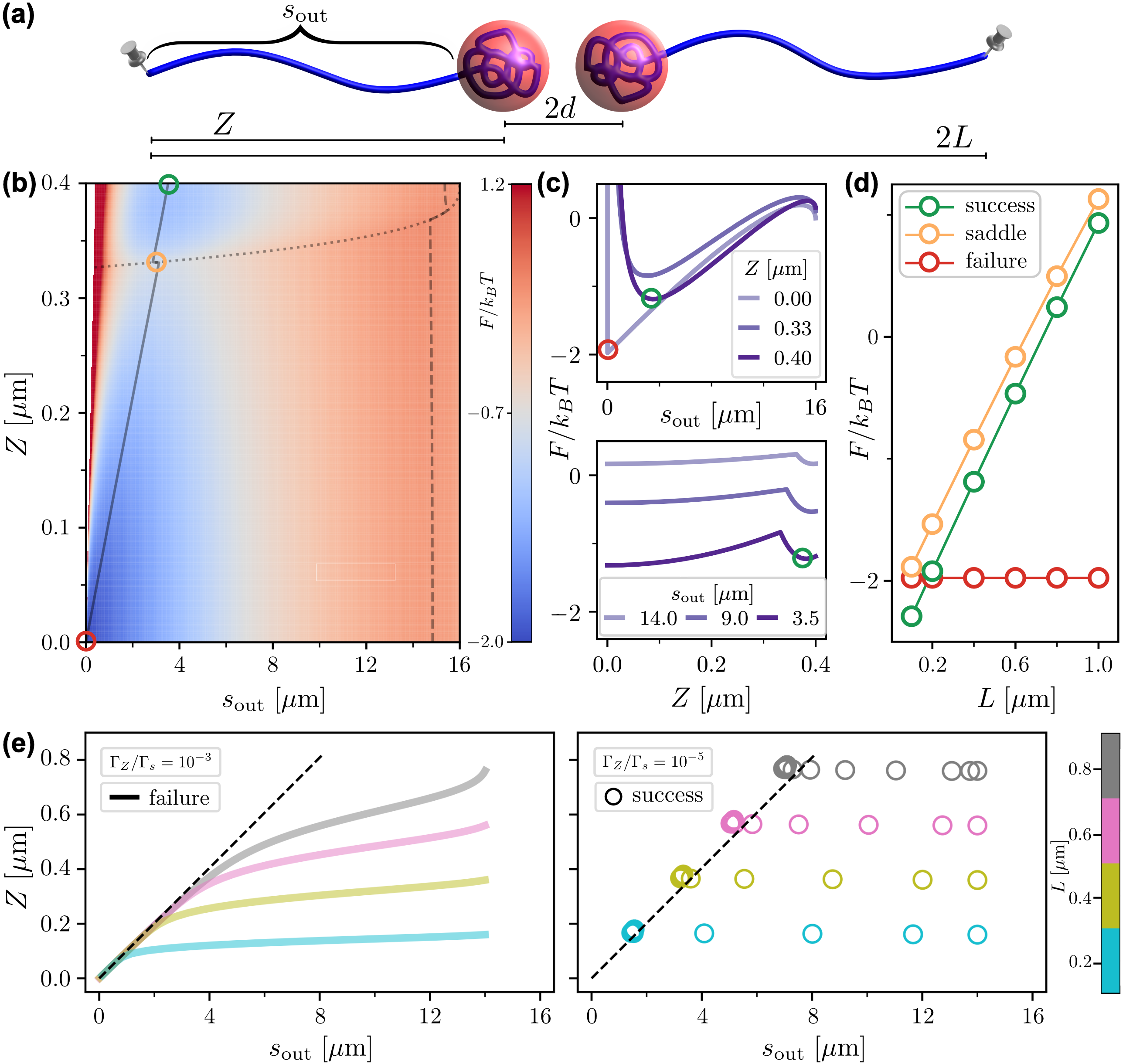} 
    \caption{
    {\bf A simple polymer–droplet model identifies distinct stationary states governing tethering outcomes.}
    {\bf a.} 
    Schematic of the simple model showing two identical broken chains (blue), with their terminal ends pinned in space, and growing condensates at their broken ends. 
    The separation between the pinned termini of the chains is $2L$, that between the left pinned terminal and the condensate center is $Z$, and the center-to-center separation between the condensates is $2d$. 
    At time $t$ after the break, a contour length $s_{\mathrm{out}}(t) \in [0, S]$ of the broken chain lies outside the condensate, while the remaining length $S - s_{\mathrm{out}}(t)$ is inside.
    {\bf b.} 
    Representative free energy landscape $F(s_{\mathrm{out}}, Z)$, where $s_{\mathrm{out}} \in [0, S]$ and $Z \in [0, L]$. 
    The conditions $\partial F / \partial s_{\mathrm{out}} = 0$ define the solid and dashed lines, corresponding respectively to a local minimum and maximum in $s_{\mathrm{out}}$ for a given $Z$. 
    The dotted line indicates $d = R$, at which the condensates touch. 
    Two stationary points, one at $Z \simeq L$ (green circle) and another at small $Z$ (red circle), correspond to the tethering success and failure states, respectively. 
    The orange circle marks a saddle point, where $F$ attains its minimum along the $d = R$ curve.
    Percentile-based scaling is used to define the colorbar for improved visualization of the basins in $F$.
    {\bf c.} 
    Free energy curves are shown as functions of $s_{\mathrm{out}}$ (top panel) and $Z$ (bottom panel) for representative values of the state variables. 
    The green and red circles are the same as those shown in subfigure b.
    {\bf d.} 
    For each $L$, $F$ is evaluated at the stationary points corresponding to tethering success and failure, as well as at a saddle point (defined as in subfigure b).
    $F\mathrm{[saddle]} - F\mathrm{[success]}$ diminishes with increasing $L$. 
    {\bf e.} 
    System evolution on $(s_{\mathrm{out}}, Z)$-space obtained by integrating eqs.~\ref{eq_dynamics_sout}--\ref{eq_dynamics_Z}, and starting from $s_{\mathrm{out}} = 14 \,{\mathrm{\mu m}}$ and a $Z = L - \Delta Z$, with $\Delta Z = 0.04 \,{\mathrm{\mu m}}$. 
    Data shown for total integration time $5 \times 10^6 \,\Delta t$ with $\Gamma_s \Delta t = 0.1\, \mathrm{\mu m/pN}$. 
    The integration time step, $\Delta t$.
    Solution to eq.~\ref{eq_quasistatic} is overlaid in black dashed line.
    $S=16 \,{\mathrm{\mu m}}$, $\alpha = 25\times 10^{-6}\,{\mathrm{\mu m}}^2$, $\nu = 13\,{\mathrm{pN/\mu m^2}}$, $\gamma = 0.2\,{\mathrm{pN/\mu m}}$, $k_BT=4.114\,{\mathrm{pN \cdot nm}}$, $P=50 \,{\mathrm{nm}}$ in {\bf b}-{\bf e}. 
    $L = 0.4 \,{\mathrm{\mu m}}$ in {\bf b}-{\bf c}. 
    }
    \label{Fig_therm_model}
\end{figure*}

In order to understand this interplay, we develop a minimal physical model of polymer--$B$ condensates, described as droplets forming on retracting semi-flexible chains.
We denote the total polymer contour length by $2S$, with termini pinned at separation $2L$ along the $\hat z$-direction (Fig.~\ref{Fig_therm_model}a).
At time $t=0$, a symmetric break generates two broken chains of equal contour length $S$, which subsequently undergo condensation from their broken ends. 
At a given $t>0$, we denote by $s_{\mathrm{out}}(t) \in [0,S]$ the contour length of the chain segment remaining outside the condensate. 
The other portion of the chain of length $S-s_{\mathrm{out}}$ forms the condensate of radius $R=\left[\alpha\left(S-s_{\mathrm{out}}\right)\right]^{1/3}$, where $\alpha$ describes the condensate volume per unit contour length.
We further define the extension $Z(t)$ of the chain as the distance between a condensate centre and its pinned terminal, and $2d(t)$ as the centre-to-centre separation between the condensates growing at the broken ends, both measured along the $\hat z$-direction.
We study this model system in the state space spanned by $s_{\mathrm{out}}$ and $Z$. 

The free energy per broken chain can be approximated as (see Methods)
\begin{eqnarray}
    F\left(s_{\mathrm{out}}, Z\right) = 
    \frac{k_B T}{P} \left[ \frac{1}{4}\frac{Z^2}{s_{\mathrm{out}} - Z} + \frac{1}{2} \frac{Z^2}{s_{\mathrm{out}}} \right]   
    - \frac{4 \pi}{3} \alpha \left[ S - s_{\mathrm{out}} \right] \nu
    + F_{\mathrm{s}}\left(s_{\mathrm{out}}, Z\right) \quad .
    \label{Eq_F}
\end{eqnarray}
Here, $P$ denotes the polymer persistence length and $\nu$ the condensation free-energy density. 
The term $F_{\mathrm{s}}(s_{\mathrm{out}}, Z)$ represents the interfacial energy due to surface tension, which can give rise to a capillary force and thereby drive condensate fusion when the condensates overlap ($d \leq R$); this capillary contribution becomes $Z$-independent when the condensates are separated (see Methods). 
In Fig.~\ref{Fig_therm_model}b, we illustrate the free-energy landscape $F\left(s_{\mathrm{out}}, Z\right)$ for a representative set of parameter values, highlighting the states associated with tethering success (green circle) and failure (red circle).
At a given extension $Z$ of the broken chain, $F$ exhibits a minimum at an intermediate value of $s_{\mathrm{out}}$, because of the competition between polymer elasticity and condensation energies (Fig.~\ref{Fig_therm_model}b, solid line; Fig.~\ref{Fig_therm_model}c, top panel).
As a function of $s_{\mathrm{out}}$, the free-energy can exhibit for large $s_{\mathrm{out}}$ a local maximum between a non-condensed and a condensed state (Fig.~\ref{Fig_therm_model}b, dashed line; Fig.~\ref{Fig_therm_model}c, top panel).
The dotted line in Fig.~\ref{Fig_therm_model}b corresponds to $d=R$ implying that two condensates touch.
Above that line the condensates undergo fusion. 
Free energy profiles for fixed $s_{\mathrm{out}}$ but varying $Z$ are shown in Fig.~\ref{Fig_therm_model}c, bottom panel.
They exhibit a minimum at large $Z$ corresponding to tethered state; non-tethered state at low $Z$ are separated by an energy barrier.
In the $(s_{\mathrm{out}}, Z)$-space (Fig.~\ref{Fig_therm_model}b), a minimum of $F$ near $Z=L$ (green circle) corresponds to successful tethering, the minimum near $Z=0$ and $s_{\mathrm{out}}=0$ (red circle) corresponds to failure.
A saddle point (orange circle) separates the two basins of attraction corresponding to these success- and failure-states in the $(s_{\mathrm{out}}, Z)$ space.
Fig.~\ref{Fig_therm_model}d shows the free energy associated with the success-state (green), failure-state (red), and saddle point (orange) as function of initial extension $L$.
For small $L$, the success-state can emerge as the global minimum, while at intermediate $L$ it persists as a metastable state protected by the energy barrier $\Delta F = F[\mathrm{saddle}] - F[\mathrm{success}]$.
This protective barrier progressively diminishes with increasing $L$, and the success-state eventually disappears for large $L$.

To investigate how the system evolves from a given initial configuration toward either of the two free-energy minima, we formulate dynamical equations for the state variables $s_{\mathrm{out}}$ and $Z$ in a linear response framework:
\begin{eqnarray}
    \frac{ds_{\mathrm{out}}}{dt} &=& - \Gamma_s \frac{\partial F}{\partial s_{\mathrm{out}}} \quad ,
    \label{eq_dynamics_sout} \\
    \frac{dZ}{dt} &=& - \Gamma_Z \frac{\partial F}{\partial Z} \quad .
    \label{eq_dynamics_Z}
\end{eqnarray}
Here, $\Gamma_Z$ is a mobility, and $\Gamma_{s}$ characterizes condensation dynamics.
The mechanical force acting on the polymer is $-\partial F/\partial Z$, while $-\partial F/\partial s_{\rm out}$
corresponds to the chemical potential imbalance between polymer segments and the condensate.
For simplicity, we do not consider a cross-coupling coefficient.
We numerically integrate eqs.~\ref{eq_dynamics_sout}~and~\ref{eq_dynamics_Z} starting from initial configurations corresponding to nascent condensates at the broken ends that have nucleated at a time $t_0$ after polymer break. 
The early polymeric retraction during this time interval leads to a value of the chain extension $Z_0 = L - \Delta Z(t_0)$, from which we begin the integration (see Extended~Data~Fig.~\ref{Extended_Data_Fig2}).
The resulting evolution trajectories are qualitatively similar to those observed in Brownian dynamics simulations (compare Figs.~\ref{Fig_simu_trajectory}e and \ref{Fig_therm_model}e). 
The system dynamics in the $(s_{\mathrm{out}}, Z)$-space is shown in Fig.~\ref{Fig_therm_model}e (right panel) for small $\Gamma_Z / \Gamma_s$ and several values of $L$. 
In this case, the trajectories relax to the corresponding success states with $s_{\mathrm{out}} > 0$ and $Z> 0$. 
In contrast, for an intermediate ratio $\Gamma_Z / \Gamma_s$, the trajectories approach the failure state, $s_{\mathrm{out}} \to 0$ and $Z \to 0$ (Fig.~\ref{Fig_therm_model}e, left panel).
These results highlight how kinetic competition between polymer retraction and condensation, as characterized via the ratio $\Gamma_{Z} / \Gamma_{s}$, can govern the tethering outcome.

In Fig.~\ref{Fig_therm_model}e, after a fast initial relaxation, the system either reaches a condensation 
equilibrium with $\partial F/ \partial s_{\mathrm{out}} = 0$, corresponding to successful tethering (see the right panel).
The condition $\partial F/ \partial s_{\mathrm{out}} = 0$ describing the balance between condensate and polymer defines a line on the $(s_{\mathrm{out}}, Z)$-space with a slope $Z/s_{\mathrm{out}} = q$, where $q$ is constant and obeys
\begin{eqnarray}
    q^2 + \frac{1}{2} \left( \frac{q}{1-q}\right)^2 =
    \frac{(8\pi/3) \alpha \nu}{k_\mathrm{B}T/P} \quad . 
    \label{eq_quasistatic}
\end{eqnarray}
This provides an analytic expression for successfully tethered states, and also captures the slowly relaxing states during failure, where $\partial F/ \partial s_{\mathrm{out}} \simeq 0$ (see the left panel).

In summary, we have shown using Brownian dynamics simulations that broken end-dependent catalysis can quickly nucleate condensates at the break-site, which can tether the broken ends. 
A complex kinetic interplay between catalysis, condensation dynamics, and polymer relaxation underlies this mechanism and can give rise to a reliable or stochastic tethering outcome. 
A simple coarse-grained polymer-droplet model captures these phenomena and shows how a fast condensation compared to the polymer relaxation dynamics can drive the system to reliable tethering. 
Nucleation is fundamentally stochastic, and fluctuations can play important role in this process. 
Our work shows that there are parameter regimes where the tethering outcome is reliable even in the presence of noise.
To study the stochastic regimes using our simple model, it requires addition of noise to eqs.~\ref{eq_dynamics_sout}--\ref{eq_dynamics_Z}, which we leave for future work. 

Our model can account for the single-DNA-molecule assay observations reported in \cite{AlbertiCell2024}, where double-strand breaks in extended $\lambda$-DNA molecules induced DNA--PARP1 co-condensation, which held the broken ends together after about a minute.
With $S = 8.25 \, \mathrm{\mu m}$, $L = 7.03 \, \mathrm{\mu m}$, and the radius of the fused condensate $\sim 0.4 \, \mathrm{\mu m}$, as estimated from the data presented in \cite{AlbertiCell2024}, and choosing $\alpha = 0.03 \, \mathrm{\mu m^2}$, $\nu = 26 \, \mathrm{pN/\mu m^2}$, and $\gamma = 2 \, \mathrm{pN/\mu m}$, we find the stationary state corresponding to the successful tethering at $s_{\mathrm{out}} = 7.54\, \mathrm{\mu m}$ and $Z = 6.82\, \mathrm{\mu m}$, which are consistent with the considered data and predicts a force of $\sim 2.3\, \mathrm{pN}$ on the broken DNA chain.
We can interpret the outcome of this experiment using our theory. 
Our work shows that localized condensation at the broken ends is required and sufficient to allow for a tethering of the broken ends by a bridging condensate. 
If condensation is not localized to the end, sustained tethering is not achieved (see Supplemental Material \cite{SM}). 
This raises the question of how reliable and fast nucleation localized at the broken ends can be achieved. 
Our work demonstrates that such fast and localized condensation is naturally generated if the broken end triggers conformational changes towards a condensation competent state. 
The molecular details of this process are still unclear and could involve conformational activation of PARP1 or cooperative interactions \cite{AlbertiCell2024}. 
To clarify such details will be a challenge for future work.

The physical picture of end-capture by end-induced condensation developed here may also be relevant in \emph{in vivo} contexts, where DNA breaks occur within the nucleus, a crowded environment rich in PARP1 and many other chromatin-associated components.
Before other repair factors are recruited to the repair site, it is important that the ends are tethered together. 
This could be achieved by a PARP1 condensate. 
Furthermore, DNA in chromatin can in general be under some tension due to active processes or capillary forces. 
Our work shows polymer tethering by a condensate can sustain forces on the order of piconewtons. 
Exploring the biophysics of DNA breakage and repair inside the cell will be challenge for future research.

\clearpage
\noindent
{\Large \bf Methods.}

\vspace{3mm}
\noindent
{\bf Simulation system.}\textemdash
We consider
\begin{eqnarray}
    u(r_{n,m}) = 4 \left[\left(\frac{\sigma}{r_{n,m}}\right)^8 - \left(\frac{\sigma}{r_{n,m}}\right)^4 \right], \quad r_{n,m} \leq \lambda , 
\end{eqnarray}
where the truncation length scale is taken to be $\lambda = 3 \times 2^{1/4}\sigma$. 
This choice maintains short-ranged interactions between the beads while avoiding numerical instabilities associated with the strong divergence of $u(r_{n,m})$ at small separations.
With $\eta$ denoting the medium viscosity, the timescale $\tau = \eta \lambda^3 / (k_B T)$ characterizes diffusion over the length scale $\lambda$.
The bead mobility is approximated using Stokes' law, $\Gamma = 1/[6\pi \eta (a/2)]$, with the choice of bead diameter $a = \lambda/3$.

We consider 
\begin{eqnarray}
    h_{\mathrm{pin}}(\bm{x}_n)  &=&   \frac{1}{2} K_{\mathrm{pin}} |\bm{X}_n - \bm{x}_n|^2, \\
    h_{\mathrm{FENE}}(r_{n,n+1}) &=& - \frac{1}{2} K_{\mathrm{FENE}} \ell^2 \ln \left[ 1 - \frac{r_{n,n+1}^2}{\ell^2} \right],
\end{eqnarray}
Here, $\bm{X}_n$ denotes a fixed position to which the bead $n$ is connected by a stiff spring with spring constant $K_{\mathrm{pin}} = 5000 \, k_BT / \lambda^2$. 
We take $(\bm{X}_{2N} - \bm{X}_1) \cdot \hat{z} = 2L$.
The FENE potential is specified via the spring constant $K_{\mathrm{FENE}} = 50\,k_BT/\lambda^2$ and the finite lengthscale $\ell = 2\lambda/3$.

We perform simulations in a rectangular parallelepiped of dimensions $24\lambda \times 24\lambda \times 384\lambda$ with periodic boundary conditions in all directions. 
This elongated geometry is chosen to reduce computational cost and is, within the studied range of $L$ and time $t \in [0, 2500\tau]$, sufficient to avoid finite-size artifacts.
Initial configurations for each value of L were generated through separate Brownian dynamics simulations of intact polymers and equilibrated, resulting in $\langle Z \rangle = L$.

To introduce a new $B$ bead at time $t>0$ during the simulations, we sample a candidate position at rate $k$:
\begin{equation}
    \bm{Y} = \bm{x}_n + (\lambda - \sigma) \bm{w}, \quad \text{for } n = N, N+1,
\end{equation}
where each Cartesian component of $\bm{w}$ is independently drawn from a uniform distribution on $(-1,1)$.
If the $\sigma$-neighborhood of $\bm{Y}$ is unoccupied by any of the $2N+M-1$ existing beads, a new bead is placed with $\bm{x}_{2N+M} = \bm{Y}$ and $\xi_{2N+M} = B$, increasing the total number of beads to 2N+M at time $t$.

\vspace{3mm}
\noindent
{\bf Free energy of the polymer segment outside condensate.}\textemdash
To estimate the entropic tension in the polymer segment outside the condensate, we adopt the phenomenological force–extension relation proposed by Marko and Siggia for semiflexible $\lambda$-DNA~\cite{SiggiaScience1994}:
\begin{eqnarray}
    f_{\mathrm{poly}} = \frac{k_B T}{P} \left[ \frac{1}{4(1 - p)^2} - \frac{1}{4} + p \right] \quad ,
    \label{eq_MarkoSiggiaforce}
\end{eqnarray}
where $p = Z / s_{\mathrm{out}}$ is the ratio of the extension $Z$ of the outside segment and its contour length $s_{\mathrm{out}}$. 
Note that here we consider $R \ll Z$ which is a good approximation for successful tethering case. 
This force is related to a free energy $F_{\mathrm{poly}}(s_{\mathrm{out}}, Z) = \int_{0}^{Z} f_{\mathrm{poly}}(s_{\mathrm{out}}, Z^\prime) dZ^\prime$ which leads to 
\begin{eqnarray}
    F_{\mathrm{poly}} \left(s_{\mathrm{out}}, Z\right) = \frac{k_B T}{P} 
    \left[ \frac{1}{4}\frac{Z^2}{s_{\mathrm{out}} - Z} + \frac{1}{2} \frac{Z^2}{s_{\mathrm{out}}} \right] \quad .
    \label{eq_Fpoly}
\end{eqnarray}

\vspace{3mm}
\noindent
{\bf The interfacial energy.}\textemdash
At any instance of time $t > 0$, the interfacial area depends on whether the condensates at the broken ends overlap or are separated (i.e., $d \leq R$ or $d > R$, respectively). 
For $d>R$, condensates are spherical with surface energy $4 \pi R^2 \gamma$ per chain, where $\gamma$ denotes the surface tension.
To estimate the surface energy for $d \leq R$, when the condensates fuse, we use a geometrical model in which two spheres of radius $R$ are separated by a centre-to-centre distance $2d$ and overlap.
In order to keep the condensate volume fixed as condensates undergo fusion, we adjust the radii to a value $\tilde{R}(s_{\mathrm{out}}, Z)$ with $\tilde{R}^3 + \frac{3}{2}\tilde{R}^2 d = 2R^3 + \frac{1}{2}d^3$, such that $\tilde{R}^3 = 2 R^3$ for $d=0$ corresponding to complete fusion, and $\tilde{R} = R$ for $d=R$ before fusion. 
This construction does correctly reproduce the limits of complete or no overlap, without capturing complexities of liquid droplet coalescence discussed in \cite{SnoeijerAnnuRevFluidMech2025}.
With this choice, we have
\begin{eqnarray}
    F_{\mathrm{s}}\left(s_{\mathrm{out}}, Z\right) = 
    \begin{cases}
    2 \pi \tilde{R} \left(\tilde{R} + d\right) \gamma \quad , & d \leq R \\
    4 \pi R^2 \gamma \quad ,                                  & d > R \quad.
    \end{cases}
    \label{Eq_Fcap}
\end{eqnarray}



\begin{acknowledgments} 
This work was funded by the Max Planck Society and supported by the Deutsche Forschungsgemeinschaft (DFG) through Physics of Life (project number 390729961), the DFG grants to N.C. (project number 419138288), and GRK 3120 (project number 542285965). 
N.C. received further supports from the DFG (SPP2191, 471025906) and the Max Planck Society.
We acknowledge support from the Max Planck Computing and Data Facility. 
We thank Jean-Fran\c{c}ois Joanny and Omar Adame-Arana for constructive comments and feedback.
\end{acknowledgments}

\clearpage

\clearpage
\noindent
{\Large \bf Extended Data.}
\renewcommand{\figurename}{Extended Data Fig.}
\renewcommand{\thefigure}{\arabic{figure}}
\setcounter{figure}{0}
\vspace{3mm}
\begin{figure*}[h!]
    \centering
    \includegraphics[width=0.95\textwidth]{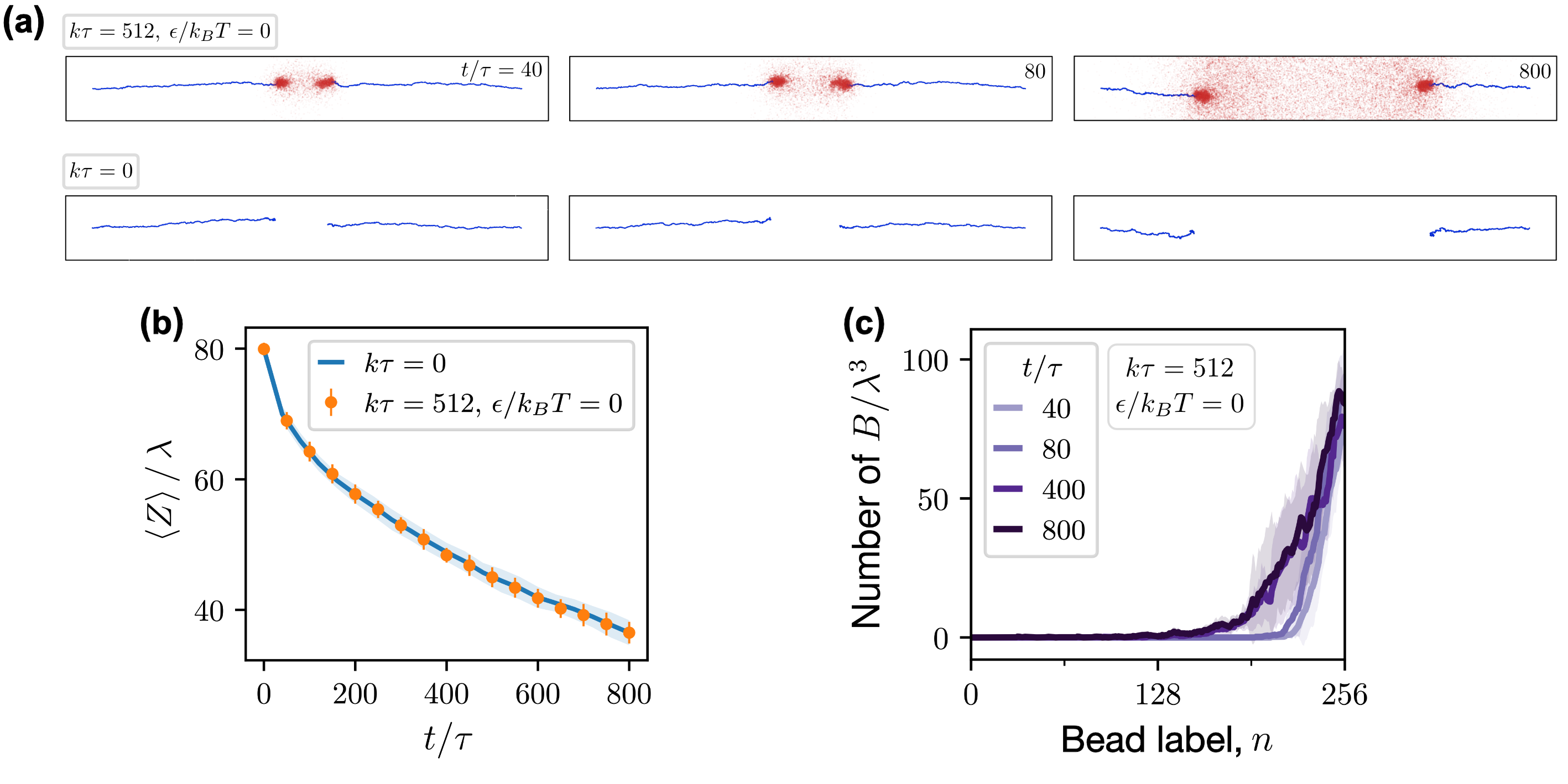} 
    \caption{
    {\bf Condensation is necessary for broken end tethering.}
    {\bf a.} 
    Additional simulation snapshots are shown with catalysis but no interaction between polymer and $B$ beads ($k\tau = 512$, $\epsilon = 0$; top panels), and compared with the case without catalysis (bottom panels). 
    $L/\lambda = 80$ and time $t$ are indicated in the top-right corners.
    {\bf b.} 
    Despite catalysis (symbols), the extension of the broken chains decreases with time, indicating tethering failure, and overlaps with the case in the absence of catalysis (lines).
    Mean $\pm$ s.d. over $40$ independent simulation runs are shown.
    {\bf c.} 
    Despite $B$ production near the broken ends due to catalysis, no polymer–$B$ condensation is observed, as evident from the absence of a plateau in the $B$-density profiles along the bead indices of the broken chain (compare with Fig.~\ref{Fig_simu_trajectory}c).
    Mean $\pm$ s.d. over $40$ independent simulation runs are shown.
    }
    \label{Extended_Data_Fig1}
\end{figure*}
\begin{figure*}[h!]
    \centering
    \includegraphics[width=0.95\textwidth]{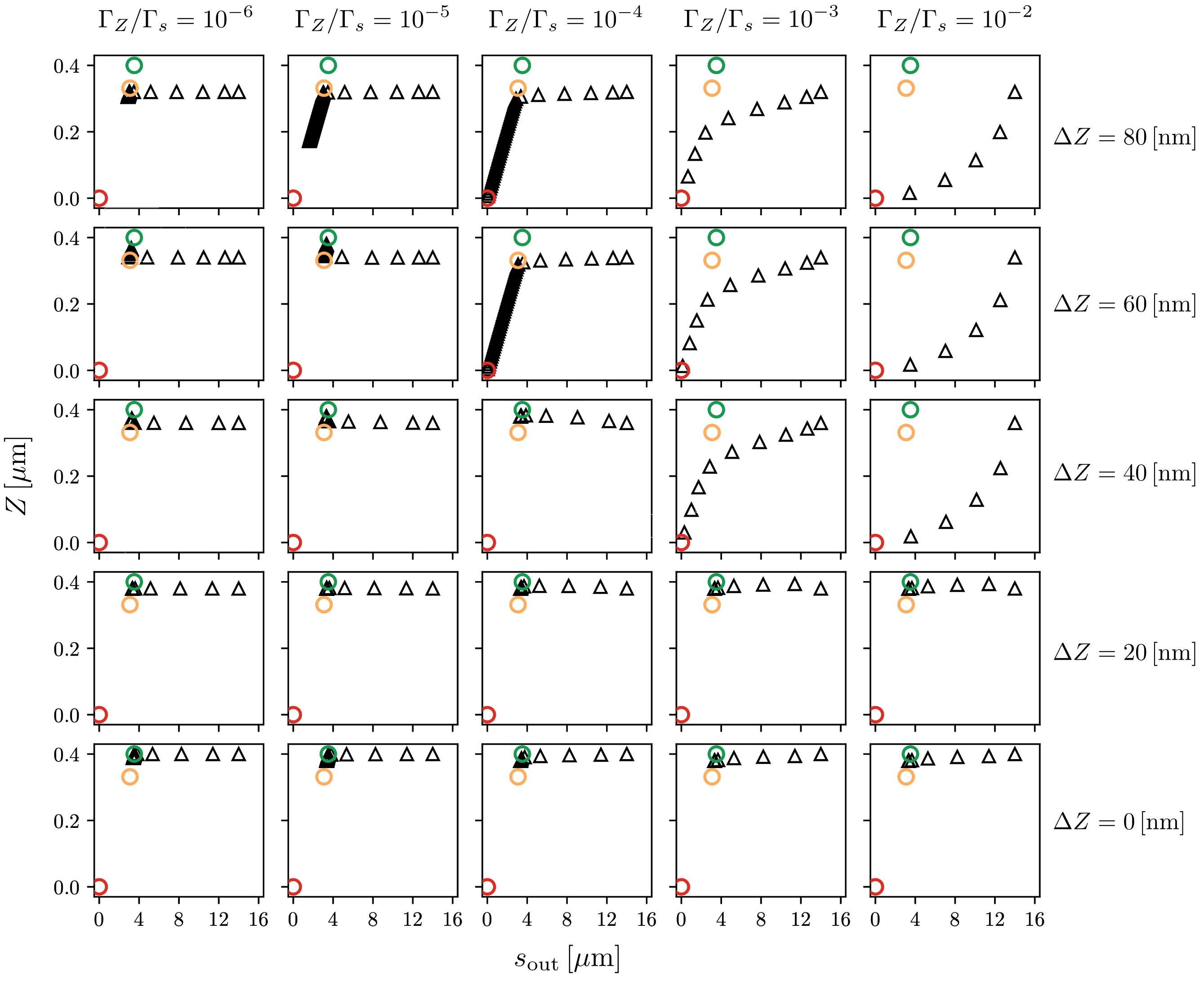} 
    \caption{
    {\bf Interplay of the kinetic competition ($\Gamma_Z/\Gamma_s$) and initial conditions in determining tethering success.}
    Evolution of the simple polymer–droplet system in the $(s_{\mathrm{out}}, Z)$ space is shown for several values of the ratio $\Gamma_Z/\Gamma_s$, starting from different initial points. 
    See Fig.~\ref{Fig_therm_model}b for the corresponding free-energy landscape; the green, red, and orange circles in each panel are the same as those in Fig.~\ref{Fig_therm_model}b. 
    As in Fig.~\ref{Fig_therm_model}e, the trajectories (black triangles) are obtained by integrating eqs.~\ref{eq_dynamics_sout}--\ref{eq_dynamics_Z} from the initial conditions $s_{\mathrm{out}} = 14\,\mathrm{\mu m}$ and $Z = L - \Delta Z$, where $\Delta Z(t_0)$ represents the early-polymeric retraction at time $t_0$ after the break, during which nascent condensates have nucleated at the broken ends. 
    The total integration time is $10^7 \Delta t$. 
    All other parameters are the same as in Fig.~\ref{Fig_therm_model}e. 
    The system evolves toward the tethering success-state for small $\Delta Z$ and/or small $\Gamma_Z/\Gamma_s$.
    }
    \label{Extended_Data_Fig2}
\end{figure*}

\end{document}